\documentclass[traditabstract]{aa} 
\usepackage{graphicx}
\usepackage{hyperref,twoopt} 
\usepackage{txfonts}
\usepackage{natbib}
\usepackage[english]{babel} 
\bibpunct{(}{)}{;}{a}{}{,} 
\begin{document}
   \title{Are solar chromospheric fibrils tracing the magnetic field?}

   \author{J. de la Cruz Rodr\'iguez\inst{1, 2} 
     \and 
     H. Socas-Navarro\inst{3, 4}
   }

   \institute{Institute for Solar Physics, Royal Swedish Academy of
     Sciences, AlbaNova University Center, SE-106 91 Stockholm, Sweden 
     \and
     Department of Astronomy, Stockholm University, AlbaNova University Center, SE-106 91 Stockholm, Sweden
     \and
     Instituto de Astrof\'\i sica de Canarias,
     Avda V\'\i a L\'actea S/N, La Laguna 38200, Tenerife, Spain
   \and
   Departamento de Astrof\'\i sica, Universidad de La Laguna, 38205, 
   La Laguna, Tenerife, Spain 
 }

\newcommand {\FeI} {\ion{Fe}{i}}
\newcommand {\CaII} {\ion{Ca}{ii}}
\newcommand {\ltau} {$\log(\tau_{5000})$}
\newcommand {\degrees} {$^{\rm o}$}

\abstract{Fibrils are thin elongated features visible in the solar
  chromosphere in and around magnetized regions. Because of their
  visual appearance, they have been traditionally considered a tracer
  of the magnetic field lines.  For the first time, we challenge that notion,
   by comparing their orientation to that of the
  magnetic field, obtained via high-resolution spectropolarimetric
  observations of \CaII \ lines. The short answer to the question
  posed in the title is that mostly yes, but not
  always. }

\authorrunning{de la Cruz R. and Socas-Navarro}
\titlerunning{Are chromospheric fibrils tracing the magnetic field?}
\maketitle
%

Narrow-band solar filtergrams in the \ion{H}{i} 6563 \AA \ (H$\alpha$) line core display an
ubiquitous pattern of fibrilar appearance covering most of the disk and 
often connecting patches of magnetic field. Although not as easily,
the same pattern is also visible in the chromospheric \CaII \ lines 
(e.g. \citealt[][]{Z74,M76}; for a recent reference see \citealt{PHZ+09}). Fibril observation requires high spatial and spectral
(particularly in \CaII) resolution, since they are very thin and
observable only in the very core of the lines. Because of their visual
appearance, which resembles magnetic field lines connecting the poles
of a magnet, it has been traditionally assumed that fibrils indeed
trace the chromospheric magnetic field. To the best of our knowledge, this
common assumption has never been verified, probably
because a proper empirical determination of the chromospheric magnetic
field is very challenging, requiring high-resolution
spectropolarimetry in chromospheric lines. Suitable
instrumentation for this purpose has only just become available. \citet{2010kuckein} were able to determine 
 the vector field in filaments (where the magnetic field is relatively strong) 
using the TIP polarimeter \citep{1999collados}.

The transverse (i.e., projected on the plane of the sky) component of
the magnetic field, which is what we are interested in for this work,
is determined solely by the observed linear polarization signals
(Stokes~$Q$ and~$U$ profiles). Unfortunately, these signals are
typically very weak and their observation presents numerous
challenges. To extract a clear signal above the noise, we
select by hand a small segment along the direction of a fibril and
average the Stokes $Q$ and $U$ profiles spatially to improve the
signal-to-noise ratio ($Q$ and $U$ are averaged separately). From the
profiles thus obtained, we can determine the azimuth of the magnetic
field on the plane of the sky.

We present here results from two different datasets acquired with two
different instruments, a Fabry-Perot interferometer and a slit
spectro-polarimeter. Owing to the nature of the instrumentation
employed, each dataset exhibits its own advantages and disadvantages
for our purposes but they complement each other well, as we discuss
below.

The first dataset is a spectro-polarimetric scan with the
Spectro-Polarimeter for INfrared and Optical Regions (SPINOR,
\citealt{SNEP+06}) of \CaII \ 8542~\AA \ at the Dunn Solar Telescope
of the National Solar Observatory/Sacramento Peak Observatory
(Sunspot, NM, USA). The observing setup and the data are described in
detail in that paper. We analyze a $(\lambda,x,y)$ cube acquired by
scanning the spectrograph slit over a 80$\times$80\arcsec\ field of
view. Seeing conditions were exceptionally good at the time and aided
by the adaptive optics system \citep{R00}, we achieved a spatial
resolution of approximately 0\farcs 6 (although this figure varies during
the scan because temporal fluctuations of the seeing). Because of its
high quality, this dataset has also been used in previous papers studying
 the chromospheric field and electric currents in sunspots
\citep{SN05d, SN05c}. High spatial resolution is very important for
the observation of fibrils, which are barely visible in typical
spectroscopic observations of more modest resolution.

The second cube was acquired  on 2008 June 6 with the Fabry-Perot
interferometer CRisp Imaging Spectro-Polarimeter
\citep[CRISP,][]{2006crisp} in full Stokes mode at the Swedish 1-m Solar
Telescope \citep[SST,][]{2003scharmer}. The \CaII~8542~\AA \ line
was sampled at 17 wavelength points across the range $\pm1.3$~\AA \  from the
core of the line, separated equidistantly by 162~m\AA. The instrumental profile of CRISP has a full width half maximum of approximately 100 m\AA \ at 8542~\AA. The images are
processed using the image resconstruction code \textsc{Momfbd}
\citep{2005noort}, according to the scheme described in
\citet{2008noort} and \citet{2010schnerr}. The polarimetric response
of the telescope is calibrated using a one~meter polarizer, mounted on the
entrance lens. Calibration images are used to fit the parameters of a
theoretical model of the telescope as in \citet{2005selbing}.

The SPINOR observations have higher spectral resolution than the CRISP ones (120~m\AA \ dominated by instrumental resolution compared to 
324~ m\AA \ dominated by spectral sampling, respectively) and slightly higher polarimetric sensitivity. The noise in the
Stokes parameters in the absence of signal (measured as the standard
deviation in the continuum away from magnetic areas) is
~4.5$\times$10$^{-4}$ and~1.3$\times$10$^{-3}$, respectively, in
units of the average quiet-Sun continuum intensity. The spatial
averaging of the profiles that we carry out in our analysis works well
in improving the magnetic sensitivity as long as the observations are
photon-noise limited. However, at some point one reaches a limit in
which the uncertainties are dominated by other factors such as the
goodness of the calibration, flat-fielding, spurious artifacts
introduced in the image reconstruction process, and so forth. In the
SPINOR case, this limit is reached at approximately
5$\times$10$^{-5}$, whereas for CRISP it is around 10$^{-4}$. The
CRISP observations, on the other hand, have much higher spatial
resolution (0\farcs2 compared to 0\farcs6) and the linear polarization reference
direction is known. It is then possible to derive the absolute azimuth
direction without any additional assumptions. For SPINOR, unfortunately,
the linear polarization reference frame was not known and we need to
resort to an {\em a posteriori} calibration of the zero azimuth using
the data themselves. For this, we used a number of penumbral
filaments visible in the sunspot photosphere and calculated the field
azimuth in those locations as explained below. A constant offset was
added to all the azimuth values and adjusted until the field
orientation matched all the filaments simultaneously. The resulting
offset uncertainty (which propagates directly into the fibril azimuth
determinations) is approximately 2\degrees.

In both cases (SPINOR and CRISP), we analyze observations of strong
magnetic fields (in the vicinity of a sunspot) to ensure that the
polarization induced in the spectral lines is produced by the Zeeman
effect. In the quiet Sun, one would have to deal with complications
due to the influence of the Hanle effect, which not only
depolarizes the light (which would be irrelevant to our study) but
also changes the relative amplitudes of Stokes~$Q$ and~$U$
\citep{2010manso2}. Most of the fibrils studied here are of a category
that is sometimes referred to as {\em superpenumbral fibrils} \citep[see e.g.][]{2004balasubramaniam} because
they originate just beyond the boundary of a sunspot penumbra.

\begin{figure*}
  \centering
   \resizebox{0.95\hsize}{!}{\includegraphics[trim=0.5cm 0.cm 0.4cm 0.7cm, clip]{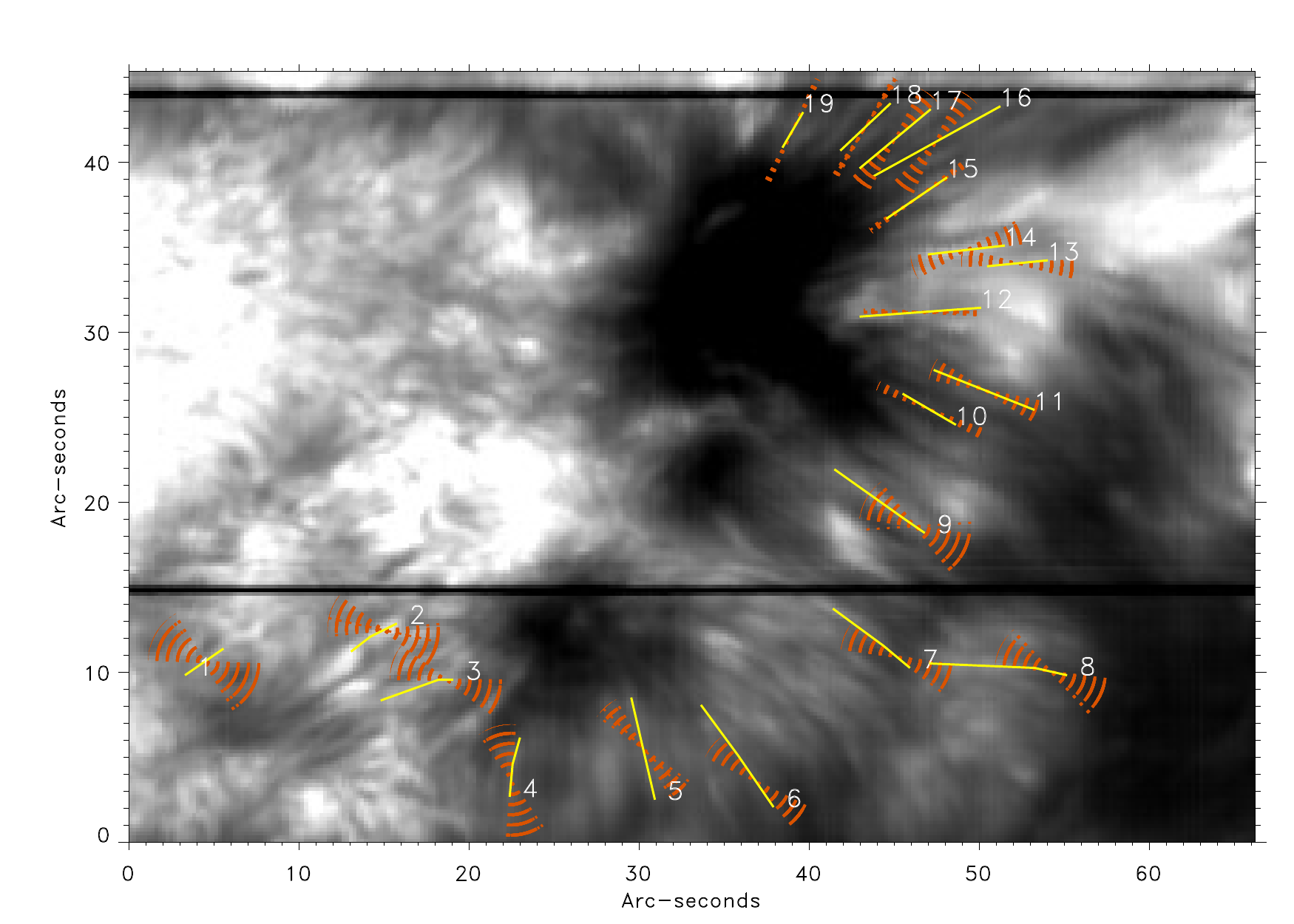}}
  \caption{Field of view observed with SPINOR in the core of \CaII \
    8542~\AA . Yellow lines: Fibrils selected for analysis. The yellow
    segments define the direction of the three-pixel wide bands used for
    Stokes $Q$ and $U$ profile averaging. Red cones: Range of
    magnetic-field azimuth compatible with the $Q$ and $U$
    profiles. The spatial sampling is 0\farcs22/pixel.
  }
  \label{fig:spinor}
\end{figure*}

Figure~\ref{fig:spinor} shows the chromospheric field of view observed
with SPINOR in a grayscale image. Superimposed on that image, 
yellow lines indicate the fibril segments that we have manually
 selected for analysis based not only on the
appearance of the fibrils but also on the presence of linear
polarization. The Stokes profiles inside the band defined by each
segment and a width of 3 pixels were averaged to produce one
low-noise set of Stokes profiles for each fibril. The azimuth
$\chi$ was then obtained using the following formula \citep{JLS89}
\begin{equation}
\label{eq:az}
\tan (2 \chi) = { \int_0^{\infty} f(\lambda) U(\lambda) \over 
  \int_0^{\infty} f(\lambda) Q(\lambda)} \, ,
\end{equation}
where $f(\lambda)$ is a bandwith selection function. In this case, we
take $f(\lambda)$ to be a rectangular function of width 300~m\AA \ centered
on the average position of the Stokes~$Q$ and~$U$ blue peak, very
close to the line core. In the CRISP case, we take a single wavelength,
where the observed linear polarization signal is maximal. The validity
of this approximation is confirmed {\em a posteriori} since it
provides azimuth values that match the orientation of the photospheric
penumbral filaments in the SPINOR dataset. We use this simpler procedure rather than full
profile inversions because it works well in determining the azimuth and
in this manner there is no need to deal with other complications
inherent to the inversion process.

 Our selection of fibrils is restricted to regions where polarization is detected after averaging. Unfortunately, the pixel-to-pixel profiles are too noisy to carry out individual measurements of the direction of the field. Only after averaging is the signal sufficiently high.  For each fibril, the azimuth was calculated using
Eq.~\ref{eq:az}. To estimate the error in the azimuth measurements, a Gaussian distribution of random values with the same sigma parameter as the standard deviation in the noise is added to $Q$ and $U$ and the
azimuth is recomputed. This procedure is repeated 100 times with
different realizations of the noise, yielding a total of 100 azimuth
values. The spread in the results obtained allows us to
estimate the uncertainties involved. The red cones drawn in the
figure on the yellow segments outline all the azimuth realizations
obtained for each fibril. Table~\ref{table:spinor} lists the values of
azimuth obtained compared to the orientation of the fibrils. The
rightmost column lists the discrepancy in units of the spread
($\sigma_{\chi}$). When this value is significantly larger than three, we
have a very high probability that the magnetic field orientation is
incompatible with that of the fibril.

\begin{table}
  \caption[]{Fibril orientation $\chi_{Fibril}$ and magnetic-field
    azimuth $\chi_{Field}$ in the chromosphere of the SPINOR dataset.}
  \label{table:spinor}
  $$ 
  \begin{array}{lllll}
    \hline
    \hline
    \noalign{\smallskip}
    Index & \chi_{Fibril} &  \chi_{Field}  \pm \sigma_\chi & 
    |\Delta \chi|/\sigma_\chi \\
    \noalign{\smallskip}
    \hline
    \noalign{\smallskip}
       1  &     34.7  &   153.4 \pm  13.0  &    9.13  \\
         2  &    216.7  &   164.0 \pm   9.0  &    5.86  \\
         3  &    195.2  &   161.5 \pm   8.0  &    4.22  \\
         4  &     80.1  &   104.4 \pm   8.4  &    2.91  \\
         5  &    103.1  &   131.6 \pm   4.7  &    6.06  \\
         6  &    125.4  &   140.1 \pm   5.1  &    2.91  \\
         7  &    142.4  &   159.1 \pm   7.3  &    2.28  \\
         8  &    175.1  &   154.8 \pm   8.2  &    2.49  \\
         9  &    144.8  &   154.6 \pm  12.0  &    0.81  \\
        10  &    149.7  &   155.9 \pm   4.3  &    1.43  \\
        11  &    158.3  &   158.5 \pm   5.3  &    0.03  \\
        12  &    184.2  &   179.2 \pm   4.5  &    1.11  \\
        13  &    185.6  &   174.2 \pm   4.0  &    2.83  \\
        14  &    186.6  &   197.7 \pm   4.6  &    2.42  \\
        15  &    214.3  &   215.8 \pm   4.1  &    0.35  \\
        16  &     28.9  &    53.7 \pm   4.8  &    5.18  \\
        17  &     39.8  &    50.4 \pm   5.1  &    2.10  \\
        18  &     43.3  &    58.1 \pm   4.2  &    3.54  \\
        19  &     59.7  &    64.1 \pm   4.0  &    1.08  \\
    \noalign{\smallskip}
    \hline
  \end{array}
  $$ 
\end{table}

We note how most fibrils are aligned with the magnetic field, although there
are a few noteworthy cases where significant misalignments occur,
well above the observational error. The most obvious are located at 
the bottom left portion of the map (fibrils 1 through 5) near the
smaller sunspot. Another interesting region is just above the large
spot, where we find fibrils perfectly aligned with the field up until
number 15, then a large misalignment in number 16, which gradually
decreases in 17 and 18 until finally number 19 is again well aligned.

\begin{figure}
  \centering
   \resizebox{\hsize}{!}{\includegraphics[trim=0.8cm 0.3cm 0.6cm 0.78cm, clip]{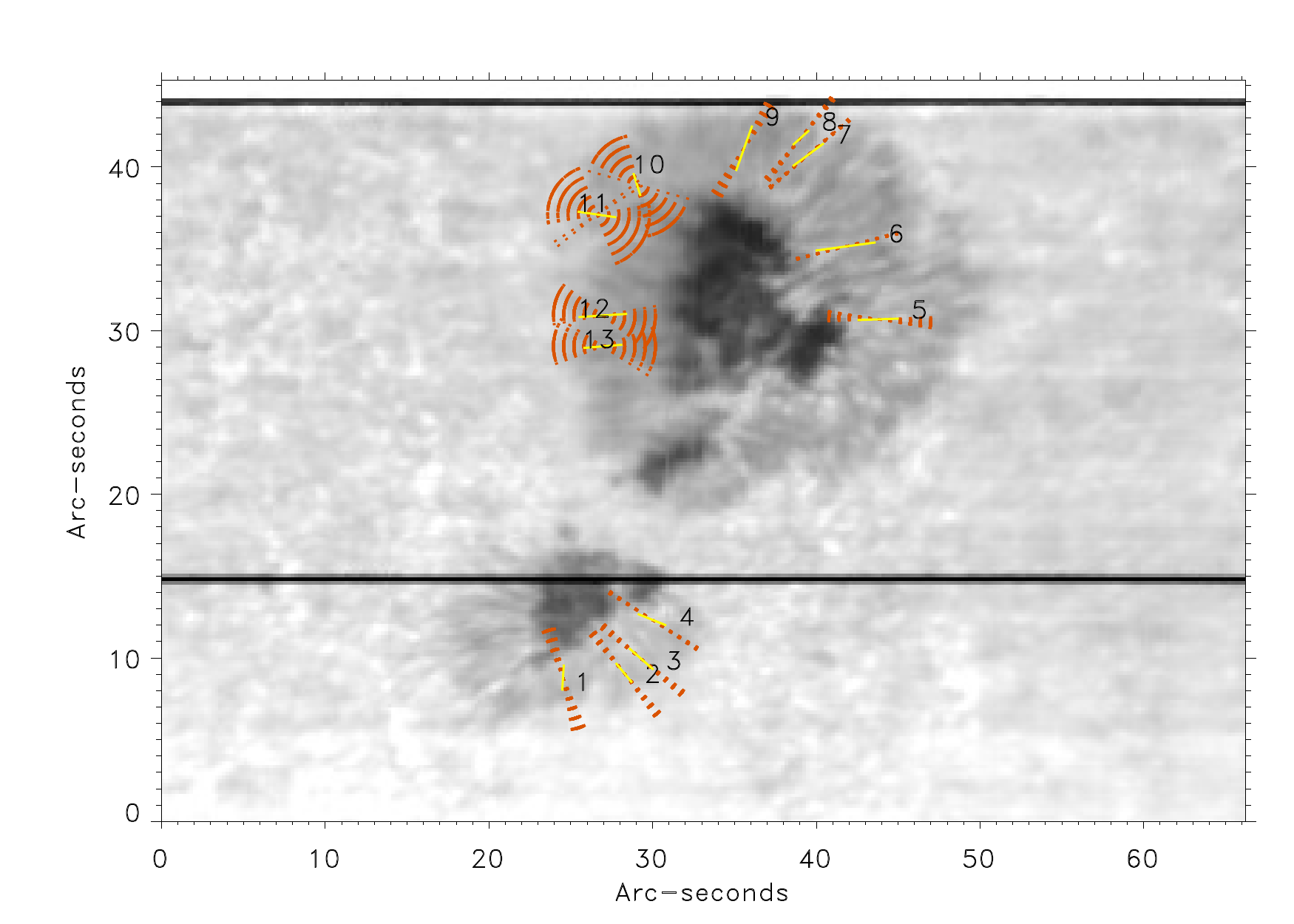}}
  \caption{Field of view observed with SPINOR in the wings of \CaII \
    8542~\AA . Yellow lines: Penumbral filaments selected for
    analysis. The yellow segments define the direction of the three-pixel
    wide bands used for Stokes $Q$ and $U$ profile averaging. Red
    cones: Range of magnetic field azimuth compatible with the $Q$ and
    $U$ profiles.  The spatial sampling is 0\farcs22/pixel. }
  \label{fig:spinorphot}
\end{figure}

For comparison, Fig.~\ref{fig:spinorphot} 
shows the same for the
photospheric penumbral filaments used to determine the absolute azimuth
reference position. The $Q$ and $U$ signals are stronger inside the
penumbra, which is why the uncertainties are smaller. The solar limb
is towards the right in the figure and therefore the transverse
component is stronger on the right-hand side of the penumbra due to
projection effects. Notice how in this figure the field is much better
aligned with the filaments. 

\begin{figure*}
  \centering
  \includegraphics[width=0.98\textwidth, trim=0 0.32cm 0 0, clip]{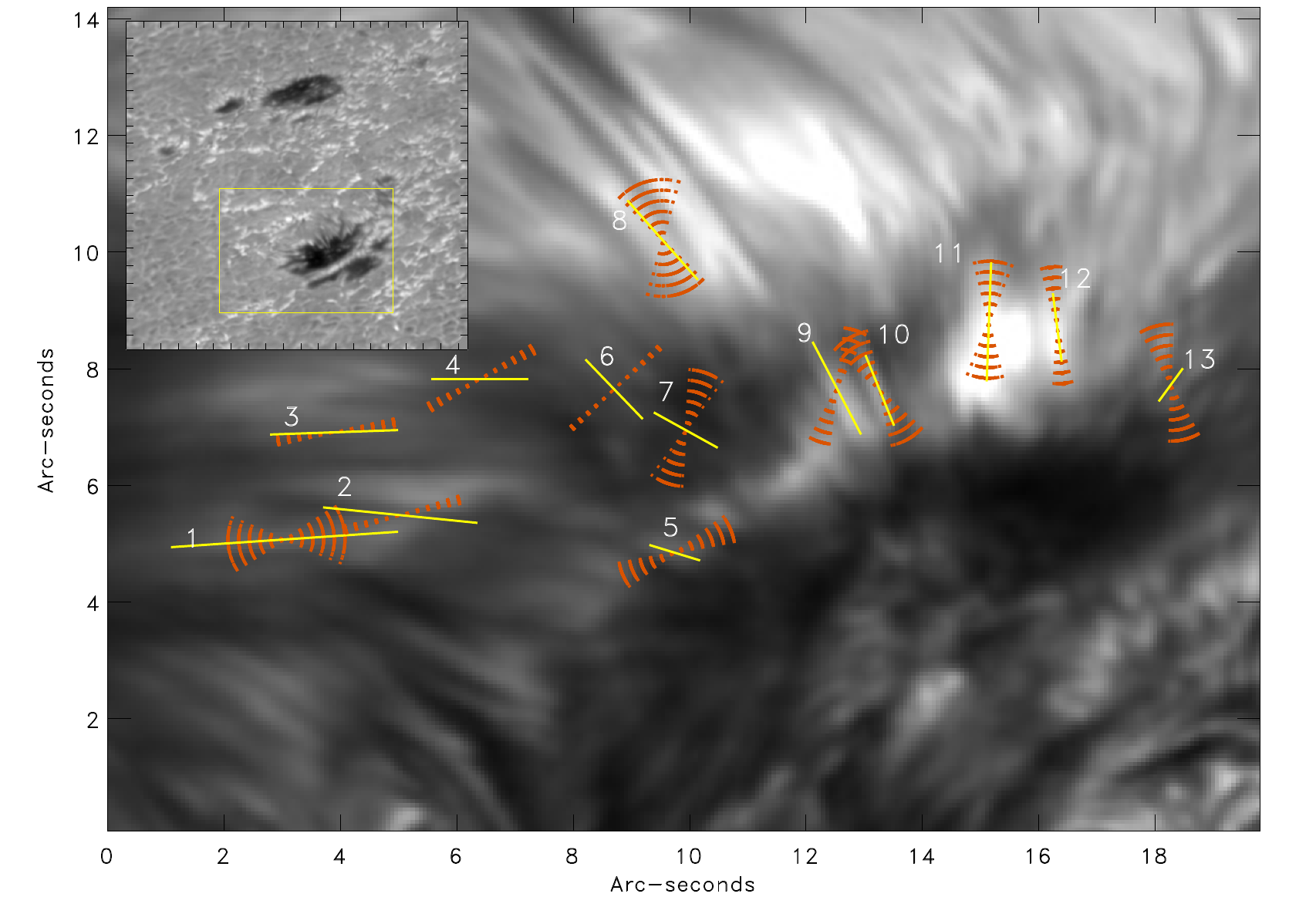}
  \caption{Field of view observed with CRISP in the wing of \CaII \
    8542~\AA . Inset: Detailed view of the line core in the region of
    interest. Yellow lines: Fibrils selected for analysis. The yellow
    segments define the direction of the three-pixel wide bands used for
    Stokes $Q$ and $U$ profile averaging. Red cones: Range of
    magnetic-field azimuth compatible with the $Q$ and $U$ profiles.
    The intensity scale has been saturated to enhance the contrast of fibrils. 
    The tick-mark separation of the inset is $2\arcsec$ covering an area of $39\times 38\arcsec$ on the surface of the Sun. 
  }
  \label{fig:crisp2}
\end{figure*}

We observe similar behavior in the CRISP data, for which the
absolute azimuth reference is known (see Fig.~\ref{fig:crisp2} and
Table~\ref{table:crisp2}). Most 
fibrils that show polarization signal have a magnetic field that is
oriented along the fibril direction, at least within the margin
allowed by the data. However, some areas (e.g., the region with
fibrils number 4, 6, 7, and 9) have a field orientation that differs
significantly from that of the fibrils. 

\begin{table}
  \caption[]{Fibril orientation $\chi_{Fibril}$ and magnetic field
    azimuth $\chi_{Field}$ in the chromosphere of the CRISP dataset.}
  \label{table:crisp2}
  $$ 
  \begin{array}{lllll}
    \hline
   	\hline
    \noalign{\smallskip}
    Index & \chi_{Fibril} &  \chi_{Field}  \pm \sigma_\chi & 
    |\Delta \chi|/\sigma_\chi \\
    \noalign{\smallskip}
    \hline
    \noalign{\smallskip}
         1  &    176.1  &   175.6 \pm  11.5  &    0.04  \\
         2  &    185.7  &   165.8 \pm   4.4  &    4.53  \\
         3  &    178.0  &   171.1 \pm   4.2  &    1.65  \\
         4  &    180.0  &   150.2 \pm   3.5  &    8.55  \\
         5  &    196.9  &   157.9 \pm   8.0  &    4.85  \\
         6  &    226.1  &   137.3 \pm   3.0  &   29.97  \\
         7  &     28.9  &   109.6 \pm   6.5  &   12.41  \\
         8  &     48.3  &    65.7 \pm  14.6  &    1.19  \\
         9  &     62.3  &   106.2 \pm   6.1  &    7.19  \\
        10  &     67.9  &    55.2 \pm   5.9  &    2.15  \\
        11  &     92.1  &    93.0 \pm  10.1  &    0.09  \\
        12  &     82.6  &    85.4 \pm   4.6  &    0.60  \\
        13  &    126.3  &    75.4 \pm   9.4  &    5.43  \\
    \noalign{\smallskip}
    \hline
  \end{array}
  $$ 
\end{table}

In the light of these results, we conclude that the widespread idea
that chromospheric fibrils are a visual proxy for the magnetic field
lines may need to be reconsidered. Here we have limited ourselves
to presenting observational evidence. An attempt to explain the
appearance of the chromospheric fibrilar pattern and the nature of
fibrils themselves is beyond the scope of the present study. We speculate that perhaps the
small difference in formation height between the \CaII \ line core
(where the fibrils are seen) and the Stokes~$Q$ and~$U$ peaks (where
the magnetic field is measured) might explain the discrepant
behavior. This would be a very surprising result because
conceiving a field topology with such strong vertical gradients in
field orientation, especially in the low-$\beta$ realm of the
chromosphere (where magnetic field pressure and tensions are more
difficult to sustain as the field dominates the dynamics of the
plasma) is very challenging. Another possibility could be that the
field changes rapidly in time and the plasma temperature structure
(which is what ultimately dictates the intensity pattern observed)
lags behind it somehow. Or perhaps the explanation is an entirely
different one. In any case, we point out that the linear polarization
signal observed in the chromosphere around sunspots weakens very
abruptly as one moves outwards from the edge of the penumbra, a 
finding that is very difficult to reconcile with the large size of the fibrilar
patterns that are seen around it, if these fibrils are indeed magnetic
field lines, because in that case the chromospheric field
strength (and the linear polarization signal) should not drop off so
abruptly as it is observed.

Although the data and the analyses presented here are reasonably well
established, our work obviously has some limitations. Given the
relevance of this subject, additional observational studies should be
conducted to confirm our refute these results. The main goal of our
Letter is to draw attention to the subject since modern
instrumentation now enables the determination of the magnetic field
vector in fibrils. A more definite and comprehensive answer to the
question raised in this paper should be something that can be
realistically expected for the near future with the existing and
upcoming tools for solar observations and their interpretation.

\begin{acknowledgements}

We are grateful to Luc Rouppe van der Voort for providing the
SST/CRISP observations. This research project has been supported by a
Marie Curie Early Stage Research Training Fellowship of the European
Communityâ Sixth Framework Programme under contract number
MEST-CT-2005-020395: The USO-SP International School for Solar Physics.\\
 Financial support by the Spanish Ministry of Science and Innovation
through project AYA2010-18029 (Solar Magnetism and Astrophysical
Spectropolarimetry) is gratefully acknowledged by HSN.

\end{acknowledgements}

\bibliographystyle{aa.bst}
\bibliography{16018}

\end{document}